\documentclass[trackchanges]{aastex7}
\usepackage{todonotes}
\usepackage{varwidth}
\usepackage{verbatim}
\usepackage{amssymb} 
\usepackage{amsmath}

\usepackage{graphicx}
\usepackage{array}

\graphicspath{{./}{figures/}}

\usepackage{float}
\usepackage{hyperref}

\begin{document}

\title{An Improved Machine Learning Approach for RFI Mitigation in FAST-SETI Survey Archival Data}

\author[[0000-0003-3977-4276]{Li-Li Zhao}\thanks{These authors contributed equally to this work.}
\affiliation{College of Computer and Information, Dezhou University, Dezhou 253023, People's Republic of China}
\email{zlili202212@163.com}  

\author[0000-0003-3977-4276]{Xiao-Hang Luan}\thanks{These authors contributed equally to this work.}
\affiliation{Institute for Frontiers in Astronomy and Astrophysics, Beijing Normal University, Beijing 102206, People's Republic of China}
\affiliation{School of Physics and Astronomy, Beijing Normal University, Beijing 100875, People's Republic of China}
\email{xhluan@mail.bnu.edu.cn} 

\author{Xin Chao}\thanks{These authors contributed equally to this work.}
\affiliation{College of Computer and Information, Dezhou University, Dezhou 253023, People's Republic of China}
\email{dzuwhf@163.com}  

\author{Yu-Chen Wang} 
\affiliation{Kavli Institute for Astronomy and Astrophysics, Peking University, Beijing 100871, People's Republic of China}
\affiliation{Department of Astronomy, School of Physics, Peking University, Beijing 100871, People's Republic of China}
\email{tjzhang@bnu.edu.cn}  

\author{Jian-Kang Li}
\affiliation{Institute for Frontiers in Astronomy and Astrophysics, Beijing Normal University, Beijing 102206, People's Republic of China}
\affiliation{School of Physics and Astronomy, Beijing Normal University, Beijing 100875, People's Republic of China}
\email{lijiankang@mail.ynu.edu.cn} 

\author[0000-0002-4683-5500]{Zhen-Zhao Tao}
\affiliation{College of Computer and Information, Dezhou University, Dezhou 253023, People's Republic of China}
\email{tzzzxc@163.com}

\author[0000-0002-3363-9965]{Tong-Jie Zhang}
\email{tjzhang@bnu.edu.cn}
\affiliation{Institute for Frontiers in Astronomy and Astrophysics, Beijing Normal University, Beijing 102206, People's Republic of China}
\affiliation{School of Physics and Astronomy, Beijing Normal University, Beijing 100875, People's Republic of China} 
\affiliation{College of Computer and Information, Dezhou University, Dezhou 253023, People's Republic of China}

\author{Hong-Feng Wang}
\affiliation{College of Computer and Information, Dezhou University, Dezhou 253023, People's Republic of China}
\email{dzuwhf@163.com}

\author{Dan Werthimer}
\affiliation{Breakthrough Listen, University of California Berkeley, Berkeley, CA 94720, USA}
\affiliation{Space Sciences Laboratory, University of California Berkeley, Berkeley, CA 94720, USA}
\email{danw@ssl.berkeley.edu}

\correspondingauthor{Tong-Jie Zhang, Hong-Feng Wang}



\begin{abstract}

The search for extraterrestrial intelligence (SETI) commensal surveys aim to scan the sky to detect technosignatures from extraterrestrial life. A major challenge in SETI is the effective mitigation of radio frequency interference (RFI), a critical step that is particularly vital for the highly sensitive Five-hundred-meter Aperture Spherical radio Telescope (FAST). While initial RFI mitigation (e.g., removal of persistent and drifting narrowband RFI) are essential, residual RFI often persists, posing significant challenges due to its complex and various nature. In this paper, we propose and apply an improved machine learning approach, the Density-Based Spatial Clustering of Applications with Noise (DBSCAN) algorithm, to identify and mitigate residual RFI in FAST-SETI commensal survey archival data from July 2019. After initial RFI mitigation, we successfully identify and remove 36977 residual RFIs (accounting for $\sim$ 77.87\%) within approximately 1.678 seconds using the DBSCAN algorithm. This result shows that we have achieved a 7.44\% higher removal rate than previous machine learning methods, along with a 24.85\% reduction in execution time. We finally find interesting candidate signals consistent with previous studies, and retain one candidate signal following further analysis. Therefore, DBSCAN algorithm can mitigate more residual RFI with higher computational efficiency while preserving the candidate signals that we are interested in.


\end{abstract}



\section{Introduction} \label{sec:intro}

Whether the Earth is the only host of life in the universe has long been a question captivating humanity. Relying on the Copernican principle and the Drake equation \citep{drake1961project}, the majority of scientists hold the view that intelligent life must exist beyond Earth (namely Extraterrestrial Intelligence, or ETI for short; \citet{2025NatAs...9...16V}). There are three primary approaches to search for extraterrestrial life: (1) direct in situ detection of biosignatures in specific environments \citep{2015Sci...347..415W}; (2) remote sensing of biological signals from the atmospheres and surfaces of exoplanets \citep{2014Sci...343..171R,2014PNAS..11112634S}; and (3) detect the technosignatures via the search for extraterrestrial intelligent, i.e. SETI \citep{1959Natur.184..844C}. From an astronomical viewpoint, the detection of biosignatures faces enormous challenges due to the constraints in detection range and observation time \citep{2005AsBio...5..706S,2014ApJ...781...54R,2016ApJ...819L..13S}. In contrast, SETI is only required to detect technological signals that are either intentionally or unintentionally emitted by an ETI. Most SETI researches are typically conducted in the radio wave band because radio signals propagate effectively through interstellar space. With advancements in radio instrumentation, the available bandwidth of the radio SETI systems has expanded to tens of GHz in recent years \citep{2018PASP..130d4502M}.

Due to broadening effects on electromagnetic signals in nature, the narrowest radio spectrum of natural astrophysical phenomena has a width of at least approximately 500 Hz \citep{1987MNRAS.225..491C}. Therefore, narrowband signals ($\sim$ Hz) are highly suitable for ETI to carry information \citep{2001ARA&A..39..511T}. When a narrowband signal is transmitted from a distant source, the relative motion between the transmitter and receiver causes a frequency shift known as Doppler drift. Therefore, the narrowband drifting signal is considered a target in SETI research.

The SETI radio observations predominantly rely on two ways: commensal sky surveys and targeted observations \citep{2001ARA&A..39..511T}. The targeted observations focus on pre-determined objects, most commonly nearby stars, while the commensal sky surveys scan large areas of sky to find potential ETI signals for further examination. For single-dish telescopes, commensal sky surveys typically employ the “drifting scan”observation strategy, leveraging the Earth's rotation to perform systematic scans across the right ascension (R.A.). The SETI commensal surveys are represented by projects such as SERENDIP program, SETI@home \citep{2001SPIE.4273..104W}, the Five-hundred-meter Aperture Spherical radio Telescope (FAST) SETI backend
\citep{zhang2020first,wang2023search}, and COSMIC \citep{2024AJ....167...35T}. In recent years, targeted SETI observations are increasingly being conducted (e.g., \citet{2013ApJ...767...94S,2017ApJ...849..104E,2017AJ....153..110G,2016ApJ...827L..22T,2018ApJ...856...31T,2016AJ....152..181H,2018ApJ...869...66H,2020AJ....160..162H,2019AJ....157..122P,2020AJ....159...86P,2020AJ....160...29S,2020PASA...37...35T,2021NatAs...5.1148S,2021AJ....161..286T,Gajjar_2021,tao2022sensitive,2023AJ....166..190T,2023AJ....166..245H,2023AJ....165..132L,Luan_2025}). Nevertheless, SETI commensal surveys can still serve as a complement to targeted observations, as they offer a few advantages: (1) they scan larger sky regions for ETI signals; (2) their observation durations are orders of magnitude longer; and (3) they are target-agnostic, thereby mitigating anthropocentric biases in target selection.

As the largest single-aperture radio telescope on Earth, FAST \citep{2000ASPC..213..523N,2011IJMPD..20..989N,2016RaSc...51.1060L,2019SCPMA..6259502J,2020RAA....20...64J} provides us with great opportunities to SETI observations \citep{nan2006five,2020RAA....20...78L}. With its L-band ($1.05-1.45$ GHz) 19-beam receiver, FAST can cover a large sky area (declinations from $-14^{\circ}.34$ to $+65^{\circ}.7$) and exhibits extremely high sensitivity ($\sim 2000\;m^{2}\; K^{-1}$). The first FAST SETI commensal survey, a drift-scan observation conducted during its commissioning in July 2019, was analyzed by \citet{zhang2020first}, who identified two groups of high-confidence ETI candidate signals, and later by \citet{wang2023search}, who identified 14 groups of ETI candidate signals from the same dataset. FAST has also conducted multiple targeted SETI observations in recent years \citep{tao2022sensitive,2023AJ....166..190T,2023AJ....166..245H,2023AJ....165..132L,Luan_2025}. In the future, FAST will conduct more SETI observations, comprising both targeted searches and commensal surveys.

The major challenge of radio SETI observations is identifying and mitigating radio frequency interference (RFI) caused by human technology. In previous analyses of FAST-SETI commensal survey data,
\citet{zhang2020first} and \citet{wang2023search} employed the Nebula platform \footnote{\url{http://setiathome.berkeley.edu/nebula}} and Hough transform method, respectively, to mitigate substantial RFI. To further remove residual RFI, both studies integrated a machine learning approach, i.e. the K-Nearest Neighbor (KNN) algorithm. Although both studies mentioned above have successfully mitigated most RFI, it is still meaningful to further improve the RFI mitigation algorithms in the following two aspects: (1) removing more RFI to reduce the work of visual inspection; (2) improving computational efficiency to accelerate data processing.

In this paper, we propose an improved unsupervised machine learning approach—the Density-Based Spatial Clustering of Applications with Noise (DBSCAN) algorithm \citep{ester1996density,schubert2017dbscan}—to replace the KNN algorithm for mitigating residual RFI. Moreover, we present the results of applying the DBSCAN algorithm to the same FAST-SETI commensal survey data analyzed in \citet{zhang2020first} and \citet{wang2023search}. In Section \ref{sec:method}, we describe the RFI removal methods proposed and used in this paper in detail. Then we briefly introduce the data and present the
results of RFI removal, candidate selection and analysis in Section \ref{sec:results}. We finally conclude this work in Section \ref{sec:conclud}.

\begin{figure}[ht!]
\centering
    \includegraphics[width=0.98\linewidth]{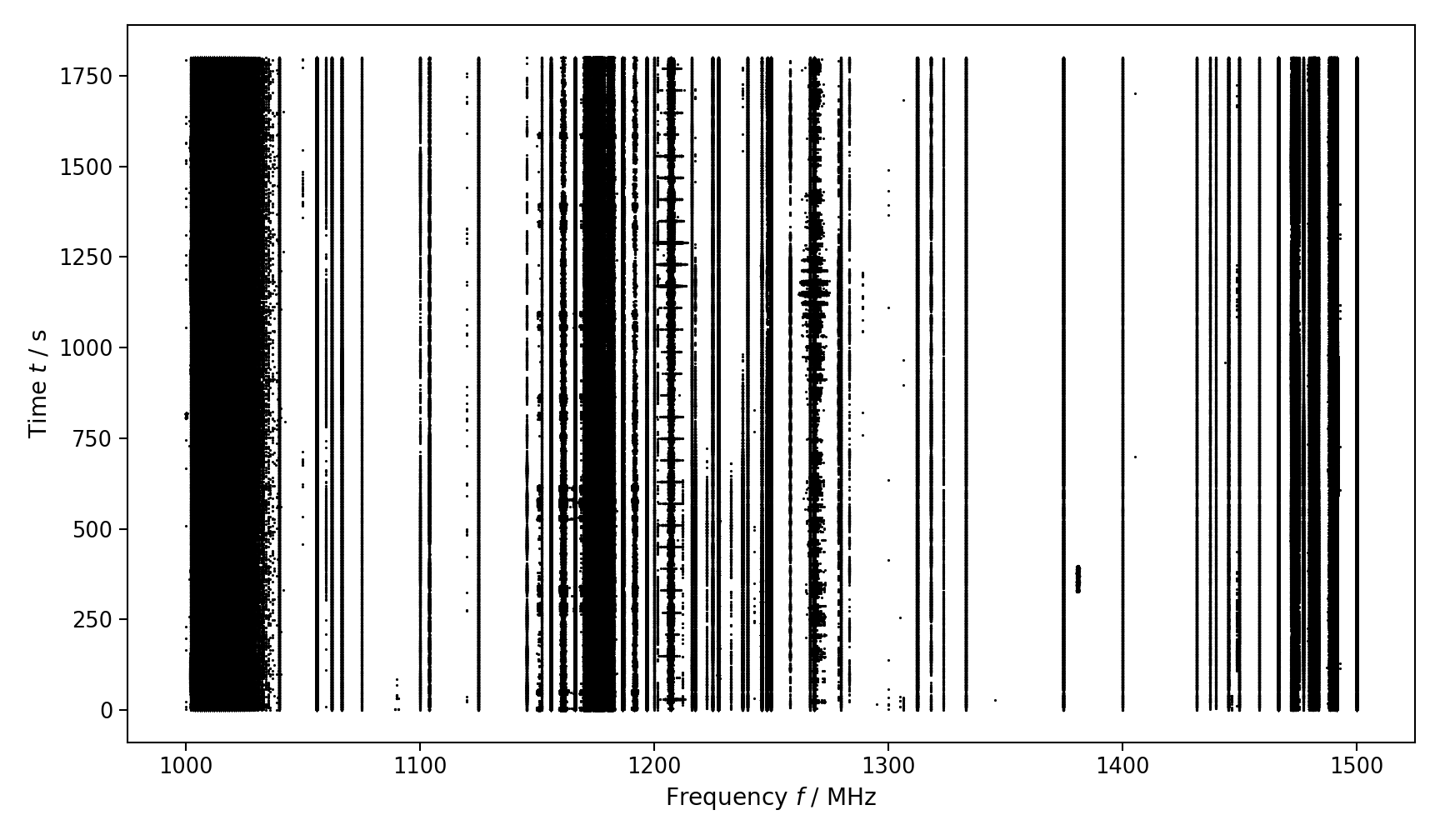}
    \caption{The waterfall (frequency-time) plot of the raw observational data (i.e., hits, marked with black dots) from the first 1800s of observation. The narrowband RFI is prominent, appearing as vertical lines; some broadband RFI (appearing as horizontal line segments) also exists. This figure is same as the Figure 1 in \citet{wang2023search}.}
    \label{fig:Original}
\end{figure}

\section{Methods for RFI Mitigation} \label{sec:method}

In the SETI commensal survey, RFI removal algorithms are primarily applied to process “hit” records as the input data. A “hit" in this paper refers to information contained in potential signal of interest that has a high signal-to-noise ratio (S/N) in its frequency channels at each moment (See Section \ref{subsec:data} and \citet{zhang2020first} for details). Each hit contains multiple observational parameters, including timestamp (UTC), frequency channel identifier, telescope pointing coordinates, and system temperature measurements, which can be used to identify RFI. As shown in Figure \ref{fig:Original}, all hits are distributed on a waterfall plot of time $t$ and frequency $f$. Most hits are RFI, while some may be interesting ETI candidates. Unlike targeted observations that use “filterbank” data with complete spectral information \citep{tao2022sensitive,2023AJ....166..190T,2023AJ....166..245H,2023AJ....165..132L,Luan_2025}, commensal surveys typically involve long-duration observations containing only hit data. Therefore, specialized software for analyzing hit data is required.

In this paper, the process of RFI removal includes three steps: (1) persistent narrowband RFI removal; (2) drifting (narrowband) RFI removal; and (3) removal of RFI using the clustering algorithm. Specifically, the third type of RFI typically includes narrowband RFI and broadband RFI, making it difficult to mitigate. We collectively designate it as “residual RFI" and use an optimized machine learning method to remove it.

\subsection{Persistent and Drifting Narrowband RFI Removal} \label{subsec:2rfi}


We remove the first two types of RFI following the methods proposed in \citet{wang2023search}. Persistent narrowband RFI typically manifests as signals within a specific frequency channel that appear across a wide expanse of the sky or endure for extended durations. Such signals are generally not of astronomical origin and appear as prominent vertical lines in Figure \ref{fig:Original}. In the SETI commensal survey observations, this persistence often translates to detection over a broad sky area as the Earth's rotation moves the telescope's pointing. For persistent narrowband RFI removal, the full frequency range (1000–1500 MHz) is divided into small bins (each $\sim$ 7.45 Hz in size). A sky angular separation threshold is then established 
as 1.5 times the distance between adjacent beam centers. When any pair of hits in a frequency bin exhibits a sky angular separation exceeding this threshold, the signal in entire frequency bin is marked as RFI (see Section 2.1 of \citet{wang2023search} for details). The characteristic of drifting (narrowband) RFI is that drifts in frequency. Such signals, which can originate from sources like satellites, moving objects, or local oscillator malfunctions, often appear as slanted or curved lines in Figure \ref{fig:Original}. For drifting (narrowband) RFI removal, the frequency-time plane is divided into a frequency window of 20 MHz and a time window of 600 seconds. Within each such window, the hits converted to binary images with pixel sizes of 0.004 MHz and 20 seconds. Line segment detection is then performed on these binary images by the $OpenCV13$\footnote{\url{https://opencv.org/}} implementation of the probabilistic Hough transform, $HoughLinesP$, which is based on the progressive probabilistic Hough transform (PPHT) proposed by \citet{matas2000robust}, and any detected line segments are extended by several pixels at each end, and all hits located within a “corridor” of less than several pixels from these extended lines are identified and removed as drifting RFI (see Section 2.2 of \citet{wang2023search} for details).

\subsection{The DBSCAN Algorithm for Residual RFI Removal} \label{subsec:dbscan}

After removing the persistent and drifting narrowband RFI as described
above, there is still a small amount of residual RFI in the data. Residual RFI typically has multiple types, and two typical examples are narrowband RFI and broadband RFI. For narrowband RFI, we cannot completely remove it using the method of Section \ref{subsec:2rfi} due to the fact that its power is sometimes below our threshold. For broadband RFI, it is easy to remove if the bandwidth is very large. However, if the bandwidth is less than several MHz, it is extremely difficult to detect using the traditional methods.

In the SETI commensal survey, the ETI signals have two limitations: they cannot last a long time (because of the pointing direction drift of the telescope, as discussed in Section \ref{subsec:candidate2}) or cover a wide frequency range (because of the commonly assumed narrowband nature). Thus, ETI signals form smaller clusters than RFI in time and frequency. So, the clustering algorithm can be used to remove residual RFI. In previous studies, \citet{zhang2020first} and \citet{wang2023search} employed the KNN algorithm to find the nearest 100 hits for each hit and calculate the mean distance. However, the KNN algorithm relies on a global density threshold to classify hits, which makes it insensitive to locally sparse RFI and challenging to identify irregularly shaped hit clusters. Moreover, the computational complexity of the KNN algorithm is O (${n}^{2}$), which becomes particularly time-consuming when dealing with massive data, potentially failing to meet real-time processing demands.


We apply an improved machine learning method, DBSCAN algorithm \citep{ester1996density}, to remove the residual RFI. DBSCAN is an unsupervised learning method, meaning it identifies inherent patterns directly from unlabeled data, which is ideal for SETI research where verified ETI labels are unavailable. This differs from the previously used KNN algorithm, which is conventionally a supervised learning algorithm but is applied for an unsupervised task by applying a threshold to the mean k-nearest neighbor distance of each hit to identify and remove large, dense clusters considered RFI \citep{wang2023search,zhang2020first}. The primary advantage of DBSCAN is using local density to define clusters, the derived unsupervised algorithm (HDBCSAN algorithm) has been successfully applied to RFI classification in other recent technosignature searches
\citep{jacobson2025anomaly}. DBSCAN is effective in separating interference signal clusters in the noise space while preserving sparsely distributed potential ETI signals by identifying arbitrarily shaped clusters in high-density regions. Due to the hits are irregular in time and frequency scales, making the clusters formed by RFI also irregular in shape. The DBSCAN algorithm is particularly well-suited to detect these irregular clusters. With the help of the robustness of the DBSCAN algorithm in noisy data, the data noise points can be effectively identified, that is, the signals that do not belong to RFI can be found.

The DBSCAN algorithm includes two core parameters: Eps and MinPts. Eps represents the radius of the neighborhood range, which is used to judge the spatial proximity between data points. MinPts represents the minimum neighborhood density threshold required for a core point. Each hit represents a black dot in Figure \ref{fig:Original}, so we describe the hits as points in the DBSCAN algorithm. In our program, the DBSCAN algorithm identifies the clusters of hits on time and frequency scales, using the Euclidean distance to define the scope of the neighborhood. The specific process is that DBSCAN scans all hits in the data set, and when the number of hits within the Eps-neighborhood of a hit (including the hit point itself) is greater than or equal to MinPts, the hit is marked as a core hit. If the number of hits within the Eps-neighborhood of a hit is less than MinPts, but the hits is in the Eps-neighborhood of a core hit, the hit is marked as a border hit. A hit is identified as a noise if it neither belongs to the core hit nor lies in the Eps-neighborhood of any core hit.

Assuming that \textit{hit1} and \textit{hit2} are different points, when the following conditions are satisfied, the sample \textit{hit1} is directly density-reachable from the sample \textit{hit2}:

\begin{enumerate}
\item The \textit{hit1} is a core hit;
\item The  \textit{hit2} lies within the Eps-neighborhood(as defined by the \citet{ester1996density}) of \textit{hit1}.
\end{enumerate}

The algorithm works by starting from each unvisited core hit, traverses all hits within its Eps-neighborhood, and adds directly the density-reachable core and border hits to the current cluster. For core hits found in the neighborhood, the algorithm recursively expands their Eps-neighborhoods as new starting points, continuously seeking and incorporating more directly density-reachable core and border hits to gradually grow the cluster. For border hits in the neighborhood, since they do not meet the density criterion of core points, their Eps-neighborhoods are not further expanded. Noise points, which do not satisfy the condition of direct density-reachability, are not assigned to any cluster. This process repeats until all core hits neighborhood have been fully traversed and no new hits can be added to any cluster. Ultimately, all the core hits and the border hits form groups that can be residual RFI clusters, and all the noisy hits form the background group.
\begin{figure}[htbp!]
    \centering
    \includegraphics[width=0.6\textwidth]{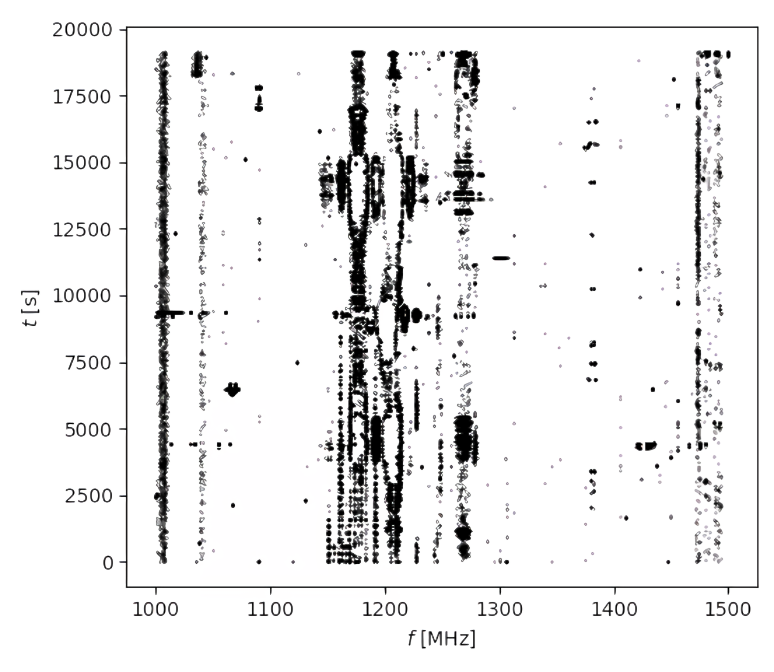}
    \caption{Waterfall plot with the hits after removing persistent and drifting narrowband RFI. The vast majority of narrowband RFI has been removed, while some broadband RFI still exists.} 
    \label{fig:example1}
\end{figure}

\section{Data and results} \label{sec:results}

\subsection{Data and Preprocessing} \label{subsec:data}

The observational data is derived from a 5-hour drift-scan observation implemented by FAST in July 2019, which is the same as \citet{zhang2020first} and \citet{wang2023search}. The data set of hits was generated from the results of SERENDIP VI \citep{archer2016commissioning,cobb2000serendip}. The SERENDIP VI is a real-time SETI spectrometer, which generates a power spectrum at each signal time, covering the 1000-1500 MHz frequency band with the frequency resolution of 3.725 Hz. The signal in the corresponding frequency channel is regarded as a “hit” when the SNR $\textgreater$ 30 \citep{wang2023search}. The “hit” information is stored in FITS format, including timestamp (UTC), frequency channel identifier, telescope pointing coordinates, and system temperature measurements, etc. The waterfall (frequency–time) plot of all hits data in the first 1800 s as shown in Figure \ref{fig:Original}. To facilitate the subsequent analysis, the data in FITS format are processed and cleaned using the Nebula system \citep{zhang2020first}. The processed FITS data are converted into a TXT format.

To verify the effectiveness of the RFI removal method, simulated ETI signals, called “birdies”, were injected into the observational data. These simulated signals comprise 20 groups with a total of 294 signals, generated and described in \citet{zhang2020first} and \citet{wang2023search}. Therefore, 47779 hits (including birdies) are obtained to verify the proposed method. The simulated signals were generated under controlled conditions emulating celestial point sources with single-frequency channel emissions, strategically positioned along FAST's observational trajectory.

\begin{figure}[ht!] 
    \centering
    \includegraphics[width=0.6\textwidth]{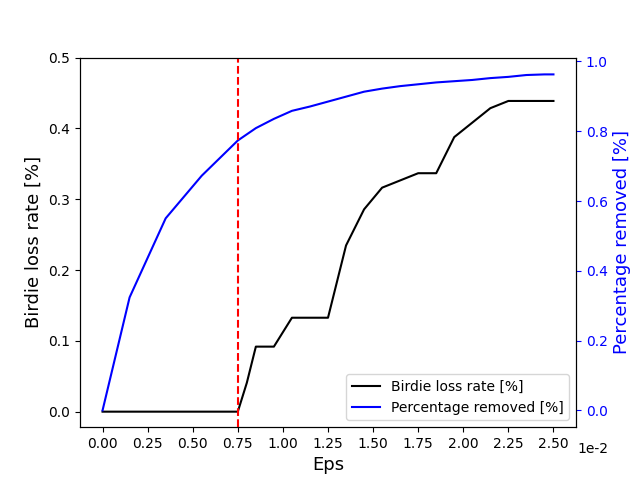} 
    \caption{Optimization of the Eps parameter for the DBSCAN-based residual RFI removal. We set the MinPts parameter is 55. The black curve (left y-axis) shows the loss rate of simulated signals(“birdies”), while the blue curve (right y-axis) shows the percentage of total residual hits removed, both as a function of the Eps value. The red vertical line indicates the selected threshold of Eps=${7.5} \times {10} ^ {-3}$, which is the optimal point that removes the largest possible percentage of RFI while ensuring the birdie loss rate remains at zero.} 
    \label{fig:birdie1} 
\end{figure}

\subsection{RFI Removal Results} \label{subsec:rfi_removal}


According to Section \ref{subsec:2rfi}, we first remove the persistent and drifting narrowband RFI, which account for approximately 99.9912\% of all data. After this removal, a total of 47779 residual hits remain, including 294 birdies. The Figure \ref{fig:example1} shows the results after removing these two types of RFI.

\begin{figure}[H] 
    \centering
    \includegraphics[width=0.98\textwidth]{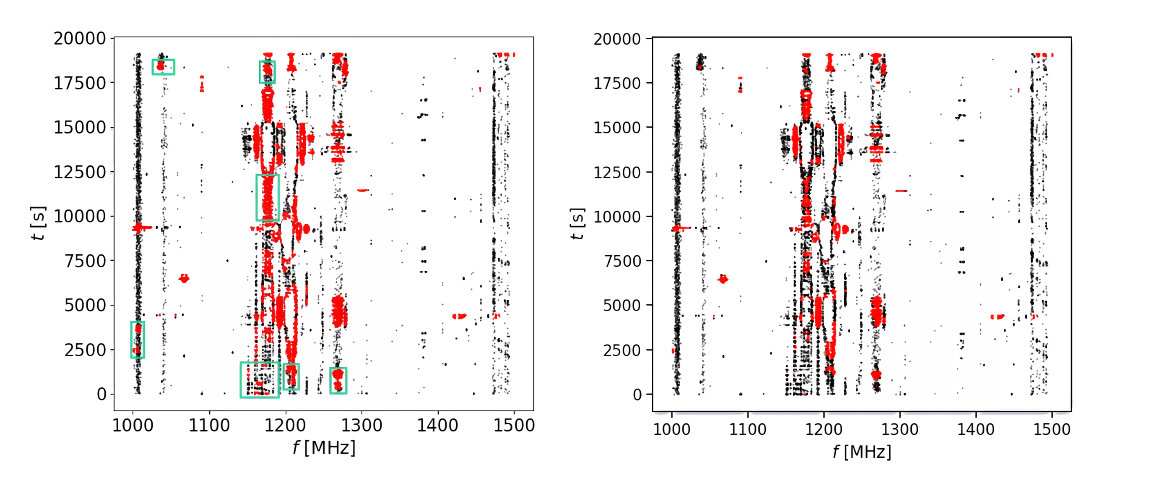} 
    \caption{Comparison result between the DBSCAN algorithm and the KNN algorithm for residual RFI mitigation. Data points marked in red represent signals identified and removed as residual RFI, while those in black are retained hits. The left panel shows the waterfall plot after applying the DBSCAN algorithm, demonstrating a residual RFI removal rate of 77.87\%. The right panel displays the waterfall plot of the same dataset using the KNN algorithm, achieving a 70.43\% removal rate \citep{wang2023search}. Notably, the green boxes in the left panel highlight RFI that DBSCAN effectively mitigation but KNN fails to identify, accounting for approximately 7.44\%.} 
    \label{fig:bothfigures} 
\end{figure}

For the residual RFI removal, we only focus on the two features of hits, namely time and frequency. Given the very wide ranges of both time and frequency, we first apply linear normalization to standardize time and frequency into the range [0,1]\footnote{The 5-hour observation duration falls within the typical observation range and excludes extreme cases in the FAST-SETI commensal surveys. For these observations, we applied the [0,1] normalization to standardize time and frequency values, making the data easier to process and compare. This ensures consistent results for different durations within a reasonable range, as minor changes in observation duration would not significantly affect the analysis when using this normalization method.}. Then, we adjust the parameters Eps and MinPts in the DBSCAN algorithm. To identify the large clusters formed by the residual RFI, the value of MinPts is set to 55\footnote{We adjusted MinPts from 35 to 95 and found that MinPts = 55 resulted in the highest residual RFI removal rate.}. The selection of Eps is determined by the simulated signal “birdies”, as shown in Figure \ref{fig:birdie1}, when the Eps is set to ${7.5} \times {10} ^ {-3}$, the loss rate of the “birdies” is 0, and the number of hits to be removed is the largest.
Based on the parameters described above, we have successfully identified and removed 36977 residual RFI (77.87\% of the residual hits) using the DBSCAN algorithm (see the left panel in Figure \ref{fig:bothfigures}). This performance is significantly higher than that of the KNN algorithm, which removed 33445 residual RFI (70.43\% of the residual hits) under identical processing conditions and the same dataset (see the right panel in Figure \ref{fig:bothfigures}, \citet{wang2023search}). After the residual RFI removal, the number of valid hits is 10508. Data processing using the DBSCAN algorithm is implemented with the scikit-learn package in Python \citep{pedregosa2011scikit}.

\begin{figure}[H] 
    \centering
    \includegraphics[width=0.7\textwidth]{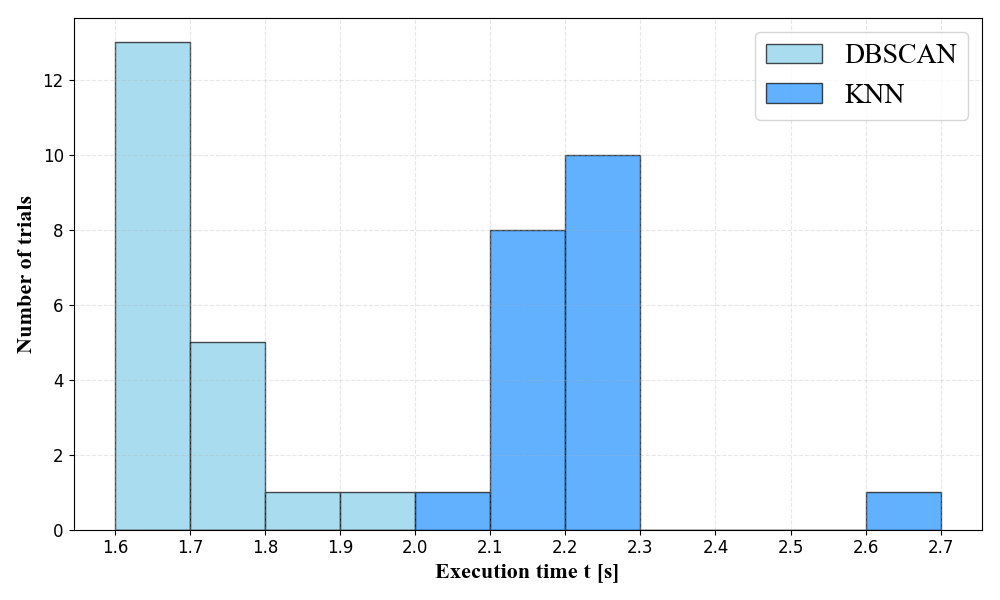} 
   \caption{Evaluation of processing speed for residual RFI removal, comparing DBSCAN(light blue) and KNN (dark blue) algorithms. The histogram shows the distribution of execution times across 20 identical runs. DBSCAN demonstrates significantly faster processing, with trials clustering in the 1.6-1.8s range, while KNN trials predominantly fall in the 2.1-2.3s range, representing an average speed improvement of approximately 24.85\%.} 
    \label{fig:speed} 
\end{figure}

Beyond the quantity of residual RFI removed, we also evaluate the processing speed during this removal process. Under identical processing conditions and using the same dataset, the DBSCAN algorithm achieves an average runtime of $\sim$1.678165 seconds for residual RFI removal, compared to $\sim$ 2.233169 seconds for the KNN algorithm, which represents a 24.85\% reduction in execution time (Figure \ref{fig:speed}). The comparison between KNN algorithm and DBSCAN algorithm for residual RFI removal is shown in Table \ref{tab:compare}.

\begin{table}
    \centering
    \begin{varwidth}{\linewidth}
    \caption{Comparison between KNN algorithm and DBSCAN algorithm for residual RFI removal}
    
    \setlength{\tabcolsep}{17pt}
    \begin{tabular}{cccc}
    \hline
    \hline
         Algorithm & Removal quantity & Removal ratio & Average execution time (s)\\
         \hline
         KNN   &  33445&  70.43\%& 2.233169 \\
         DBSCAN&  36977&  77.87\%& 1.678165\\
         \hline
    \end{tabular}
    
    \label{tab:compare}

    \footnotesize
    \textit{Notes.} \\
     Under identical processing conditions, we use the scikit-learn package in Python to implement both algorithms on the same dataset for residual RFI removal.\\
\end{varwidth}    
\end{table}

\subsection{Candidate Signal Selection} \label{subsec:candidate1}

\begin{figure}[htbp!] 
    \centering
    \includegraphics[width=0.6\textwidth]{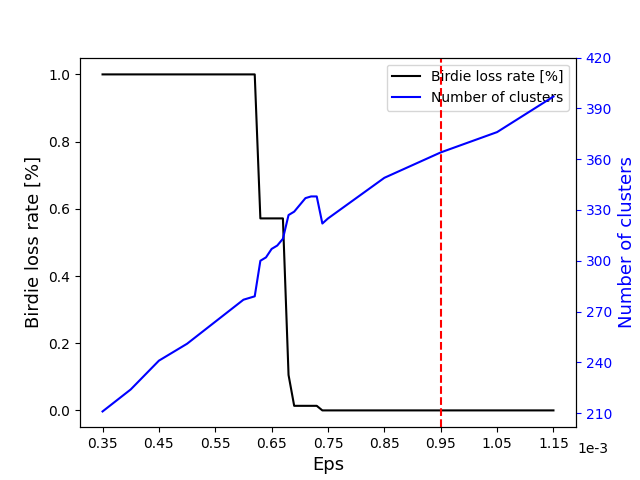} 
    \caption{Selection of the DBSCAN Eps parameter for identifying candidate ETI signals (MinPts = 5). The plot shows the birdie loss rate [\%] (black line, left y-axis) and the number of clusters (blue line, right y-axis) as a function of Eps. The chosen Eps of ${9.5} \times {10} ^ {-4}$(indicated by the red dashed vertical line) ensures a 0\% Birdie loss rate (guaranteeing all simulated signals are retained) while selecting ETI candidates from hits remaining after RFI removal.} 
    \label{fig:birdie2} 
\end{figure}

After mitigating the RFI, we select candidate ETI signals from the remaining 10508 hits. Characteristically, such candidate signals do not form large clusters in time-frequency scales; instead, they are narrowband in frequency and have a duration no longer than the telescope’s observation time for a single sky point during the drift scan. We also employ the DBSCAN algorithm to identify these small clusters, setting MinPts to 5 (much less than the threshold used to remove residual RFI), as follows the selection of MinPts for ETI candidate signal search by \citet{wang2023search} and \citet{zhang2020first}. With all simulated signals guaranteed to remain, the Eps is set to ${9.5} \times {10} ^ {-4}$ (Figure \ref{fig:birdie2}). After the candidate signal selection using DBSCAN algorithm, 364 clusters are identified, including 20 birdie clusters. 

Following the Section 3.3 of \citet{wang2023search}, the candidate clusters are further selected based on the following conditions:

\begin{enumerate}
\item The maximum sky separation of hits within a candidate cluster is less than 1.5 times the distance between the centers of adjacent beams;

\item The bandwidth of the candidate cluster is less than 500 Hz and the duration is less than 100 s.
\end{enumerate}

\begin{figure}[htbp!] 
    \centering
    \includegraphics[width=0.6\textwidth]{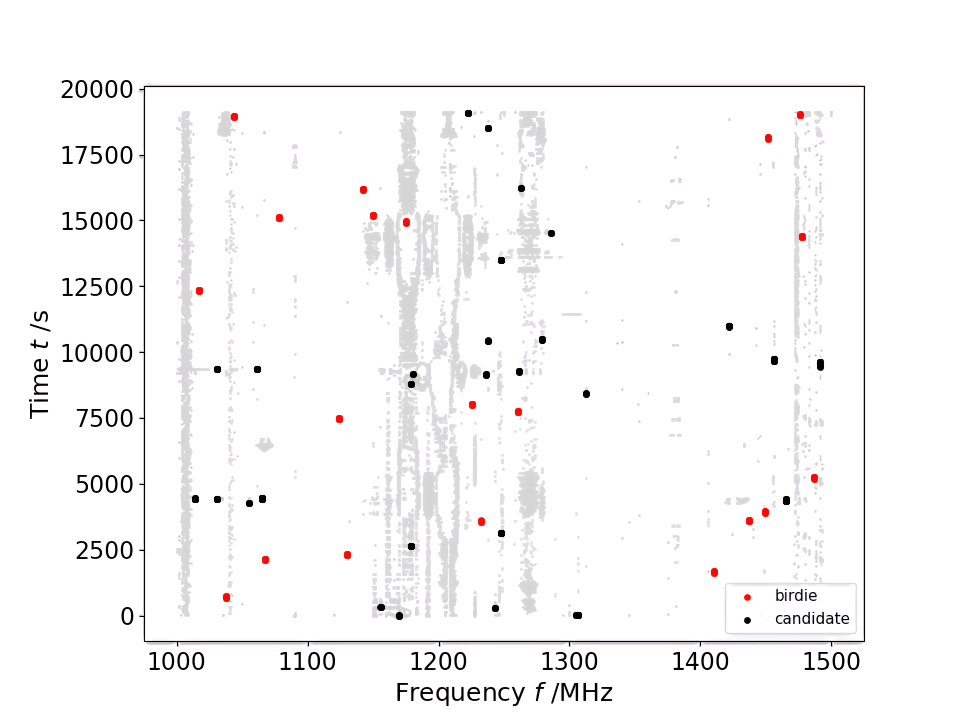} 
    \caption{Selected ETI candidates (33, black dots) and identified birdies (20, red dots) in the frequency ($f$)-time ($t$) plane, with other filtered hits shown in grey. The methodology's effectiveness is demonstrated by the recovery of all 20 “birdies” and a final candidate count of 33 (e.g., 83 by \citet{zhang2020first}; 31 by \citet{wang2023search}). } 
    \label{fig:final} 
\end{figure}

Through the above conditions, all 20 birdie clusters are marked which is consistent with the real number of birdie groups mentioned in Section \ref{subsec:data}, and 33 candidate clusters are selected, as shown in Figure \ref{fig:final}. The final cluster count demonstrates a significant reduction compared to the 83 candidates reported by \citet{zhang2020first}, and similar to (overlap substantially ) the 31 candidate clusters found by \citet{wang2023search}. This result corroborates the our methodological validity.

\begin{figure}[htbp!] 
    \centering
    \includegraphics[width=0.98\textwidth]{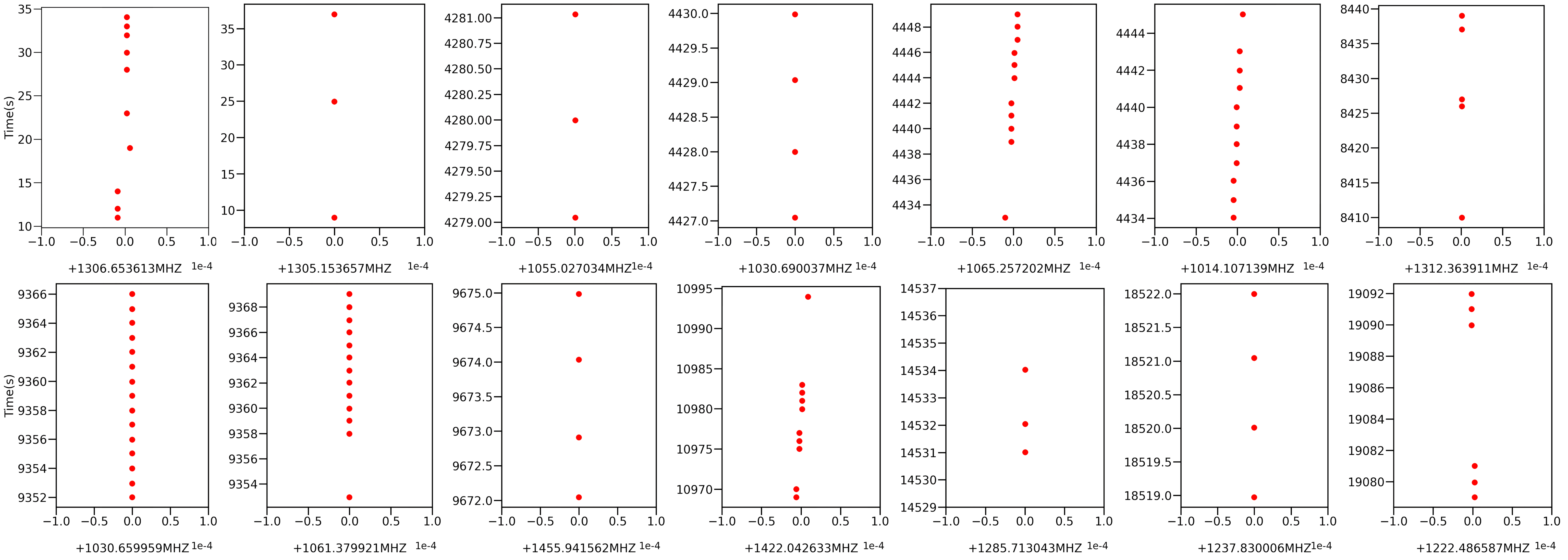} 
    \caption{Visualizations of 14 interesting candidate groups.
    These candidates were initially identified based on ETI signal characteristics and subsequently passed visual inspection, a process that confirms they do not exhibit obvious RFI features. These 14 groups are part of 33 such candidates selected in this work and are equal to those previously found by \citet{wang2023search}.} 
    \label{14candidate} 
\end{figure}

To improve the accuracy of the screening results, we use the visual inspection (see \citet{wang2023search} for details) to check whether the 33 candidate clusters above have obvious RFI characteristics. According to \citet{wang2023search} and \citet{zhang2020first}, since some selected clusters may actually be very close to other hits removed as RFI, these clusters need to be eliminated and regarded as parts of the RFI that were missed in previous steps. Specifically, \citet{wang2023search} extracted and checked the raw data within distances of 0.1 MHz, 1000 s to the candidate groups (all previously identified RFIs are marked rather than removed in the raw data), and found 14 promising groups that do not seem to be part of large RFI clusters. Similarly, using this visual inspection method, we also select 14 interesting candidate groups (consistent with those identified by \citet{wang2023search}) from the 33 candidate clusters, as shown in Figure \ref{14candidate}. More detailed information of these candidate groups is shown in Table \ref{tab:14groups}. These results show that the ETI signal is less likely to be misidentified as RFI after the removal of RFI using the DBSCAN algorithm due to the reduced amount of data. In addition, the method can not only remove more RFI, but also effectively improve the work efficiency of candidate signal selection. 

\subsection{Candidate Signal Analysis} \label{subsec:candidate2}


The 14 interesting candidate groups described above are mainly results obtained by machine learning algorithms, and further analysis still needs to be conducted. First, since the effective band of the FAST L-band receiver is 1050–1450 MHz \citep{2011IJMPD..20..989N}, we remove four candidate groups within the invalid bands (50 MHz wide each) at both ends of this range. Second, according to environmental monitoring of RFI at the FAST site, there are two sources of RFI in the observation band: civil aviation and navigation satellites \citep{2021RAA....21...18W}. There are five groups located in the civil aviation band (1030-1140 MHz) and one group located in the navigation satellite band (1176.45 $\pm$ 1.023 MHz, 1207.14 $\pm$ 2.046 MHz, 1227.6 $\pm$ 10 MHz, 1246.0-1256.5 MHz, 1268.52 $\pm$ 10.23 MHz, and 1381.05 $\pm$ 1.023 MHz). In addition, the frequency of Group No. 12 is located on the edge of one of the navigation satellite bands. Since satellite frequencies usually drift, we also consider this frequency to be approximately included in the navigation satellite band. Therefore, in order to eliminate the impact of this interference, we exclude seven candidate groups located within these frequency bands.

Apart from the effective frequency band of FAST, it is also necessary to consider the possible duration of an ETI signal. In the FAST drifting scan mode, a signal from an extraterrestrial origin passes through the beam due to the Earth's rotation. The duration of the signal depends on the half-power beamwidth (HPBW) of the telescope and its declination. The angular velocity of Earth's rotation is $15^\circ/h$, (or $0.25'/s$). Since our observation sources are located near the equator (i.e., Dec $\approx$ 0). Therefore, for a narrowband ETI signal, the maximum duration it can appear within a single beam is given by:
\begin{equation}
\boldsymbol{t = \frac{HPBW}{0.25'/s \times \cos\delta}}
\end{equation}
where $\delta$ is the declination. Since the declination of the signal source is approximately 0, $\cos\delta=1$. The HPBW decreases with the increase of frequency \citep{2020RAA....20...64J}. Within the effective frequency band, the lowest frequency corresponds to the maximum HPBW, which in turn corresponds to the longest duration. Taking the central beam (M01) as an example, when the frequency f=1060 MHz (the minimum frequency in Table 2 of \citet{2020RAA....20...64J}), the HPBW is $3.44'$; in this case, $t_{max}=13.76s$. That is, the maximum duration of an ETI signal within a single beam is 13.76 s; if a signal persists longer than this duration, it should be identified as RFI. Therefore, we exclude four candidate groups in Table \ref{tab:14groups}, as their durations are longer than the maximum value ($t_{max}$).

In conclusion, only Candidate Group No. 14 from Table \ref{tab:14groups} remain, and it cannot be ruled out as RFI based on current information. This candidate group has a short duration ($\sim 2.8$ s) and show no frequency drift. In fact, due to the Doppler effect, the frequency of a signal from an extraterrestrial origin will drift as the observation duration increases \citep{tao2022sensitive}. Therefore, we will conduct longer-duration observations to strive for a more comprehensive judgment in the future.

\begin{table*}[!ht]
\caption{Detailed Information of 14 Interesting Candidate Groups}
\centering
\setlength{\tabcolsep}{0.16cm}
\begin{tabular}{lccccccc}
\hline \hline Group & Beam & Starting time & Duration & Starting freq & Drift rate & Starting position & Off position\\
No. & No. & (JD) & (s) & (MHz) & (Hz $s^{-1})$ & (J2000 R.A. J2000 Decl.) & (J2000 R.A. J2000 Decl.) \\
\hline
1 & 08 & 2458682.209282 & 22.999996 & 1306.653604 & 0.4859 & 19:49:28.54 00:39:59.75 & 19:49:51.62 00:40:05.25 \\
2 & 08 & 2458682.209259 & 27.800010 & 1305.153657 & 0.0000 & 19:49:26.70 00:40:00.84 & 19:49:54.43 00:39:59.44 \\
3 & 15 & 2458682.258681 & 2.600082 & 1055.027034 & 0.0000 & 21:00:22.51 00:44:04.92 & 21:00:24.50 00:44:04.92 \\
4 & 14 & 2458682.260396 & 2.649986 & 1030.690037 & 0.0000 & 21:01:44.77 00:38:11.80 & 21:01:47.31 00:38:10.99 \\
5 & 14 & 2458682.260465 & 15.599996 & 1065.257192 & 0.9552 & 21:01:50.78 00:38:12.68 & 21:02:06.43 00:38:13.76 \\
6 & 15 & 2458682.260477 & 10.799982 & 1014.107134 & 1.0348 & 21:02:03.19 00:43:09.82 & 21:02:14.01 00:43:12.33 \\
7 & 19 & 2458682.306493 & 29.000004 & 1312.363911 & 0.0000 & 22:09:39.21 00:42:08.92 & 22:10:08.22 00:42:09.18 \\
8 & 04 & 2458682.317398 & 13.800006 & 1030.659959 & 0.0000 & 22:24:38.18 00:32:02.68 & 22:24:51.98 00:32:03.23 \\
9 & 04 & 2458682.317407 & 15.999994 & 1061.379921 & 0.0000 & 22:24:38.95 00:32:03.99 & 22:24:55.01 00:32:02.61 \\
10 & 18 & 2458682.321102 & 2.690026 & 1455.941562 & 0.0000 & 22:30:33.36 00:46:57.22 & 22:30:35.86 00:46:56.51 \\
11 & 04 & 2458682.336111 & 24.799986 & 1422.042627 & 0.6009 & 22:51:39.24 00:31:47.89 & 22:52:04.13 00:31:48.62 \\
12 & 10 & 2458682.423495 & 2.799985 & 1237.830006 & 0.0000 & 00:58:24.52 00:26:42.83 & 00:58:27.58 00:26:45.76 \\
13 & 01 & 2458682.429685 & 12.900001 & 1222.486589 & $-0.2888$ & 01:06:58.42 00:36:48.51 & 01:07:10.52 00:36:49.59 \\
14 & 11 & 2458682.377340 & 2.800025 & 1285.713043 & 0.0000 & 23:51:23.12 00:26:40.57 & 23:51:25.90 00:26:40.50 \\
\hline
\end{tabular}
\label{tab:14groups}
\end{table*}

\section{Discussion} \label{sec:discuss}

There are many clustering algorithms available, with HDBSCAN(Hierarchical Density-Based Spatial Clustering of Applications with Noise) \citep{campello2013density}  and OPTICS (Ordering Points to Identify the Clustering Structure) \citep{ankerst1999optics} being notable variants and improvements of the DBSCAN algorithm. These advanced algorithms demonstrate effectiveness in handling variable density clusters and noisy data, with HDBSCAN, for example, successfully employed in recent RFI clustering studies\citep{jacobson2025anomaly}. We perform comparative tests to ensure our selection of DBSCAN for residual RFI removal is optimal for our specific ETI signal search pipeline.

HDBSCAN is an extension of DBSCAN that improves performance by adding hierarchical clustering and the ability to handle variable density clusters. We adjust three primary parameters for HDBSCAN: min\_cluster\_size (minimum points to form a cluster, range 2–101), min\_samples (minimum points for a core point, range 2–31), and cluster\_selection\_epsilon (density threshold for cluster selection, range 0.0–0.05). The best residual RFI removal rate achieved is 80.6\% with the parameter combination  min\_cluster\_size = 87, min\_samples = 19, and cluster\_selection\_epsilon = 0.0, but this configuration resulted in a birdies loss rate of 13.27\%.

OPTICS is another density-based clustering algorithm that, unlike DBSCAN, allows for variable density clusters and provides a reachability plot to better visualize the structure of the data. The key parameters tuned for OPTICS are: min\_cluster\_size (minimum points to form a cluster, range 2–90), min\_samples (minimum points for a core point, range 2–40), and xi (the density threshold for cluster extraction, range 0.0–0.05).The optimal parameter combination is min\_cluster\_size=32, min\_samples=2, and xi=0.0000. This configuration achieves a birdies loss rate of 0\%, but only provides a residual RFI removal rate of 53.65\%.


For our relatively small scale dataset, our primary criterion for algorithm selection is minimizing birdies signal loss, with the objective of achieving a 0\% loss rate while maximizing residual RFI removal. Despite exploring the parameter space for HDBSCAN and OPTICS, neither algorithm outperformed DBSCAN in this key area. HDBSCAN led to unacceptable signal loss, while OPTICS resulted in a significantly lower RFI removal rate. Moreover, compared to DBSCAN, both HDBSCAN and OPTICS involve more complex parameter tuning and have higher computational complexity. Considering these factors, DBSCAN is selected as the core algorithm for the residual RFI removal step in our pipeline.

\section{Conclusion} \label{sec:conclud}

In this paper, we propose an improved machine learning approach (i.e. DBSCAN algorithm) for RFI mitigation based on the 5 hr data of FAST commissioning drift-scan survey in July 2019 \citep{zhang2020first}. Apart from mitigating the persistent and drifting narrowband RFI, we apply the DBSCAN algorithm to remove the residual RFI, which is usually challenging to mitigate. The results show that the DBSCAN algorithm can successfully identify and remove 36977 residual RFIs (accounting for $\sim$ 77.87\%), whereas the KNN algorithm achieves only 70.43\% removal rate \citep{wang2023search}. Under identical processing conditions and using the same dataset, DBSCAN also demonstrates a 24.85\% reduction in execution time compared to KNN. Finally, through the candidate signal selection, we find 14 interesting candidate groups, consistent with the results of \citet{wang2023search}. Following further analysis, one of these candidate groups is retained. These findings further verify the effectiveness of the DBSCAN algorithm.


In the future, the DBSCAN algorithm will continue to be used to remove RFI and search for more valuable candidate ETI signals in the SETI commensal survey. As the FAST-SETI dataset expands, we aim to develop a comprehensive end-to-end system for automated RFI mitigation and candidate ETI signal selection, enabling real-time processing of the observational data. Furthermore, we plan to ensemble more deep learning techniques into this framework to replace manual computation, thereby enhancing processing speed and accuracy while minimizing human intervention.

We sincerely appreciate the referee's suggestions, which helped us greatly improve our manuscript. This work was supported by National Key R\&D Program of China No.2024YFA1611804, the Shandong Provincial Natural Science Foundation (ZR2024QA180), the China Manned Space Program with grant No.CMS-CSST-2025-A01, Dezhou City Project No.2022dzkj097, Horizontal Project No.HXKT2023021, School-Level Project No.023XKZX017 and No.2022xjrc211. This work was finished on the servers from FAST Data Center in Dezhou University. This work made use of the data from FAST (Five-hundred-meter Aperture Spherical radio Telescope). FAST is a Chinese national mega-science facility, operated by National Astronomical Observatories, Chinese Academy of Sciences.

\bibliography{sample7}{}
\bibliographystyle{aasjournalv7}



\end{document}